\documentclass[acmsmall]{acmart}

\AtBeginDocument{%
  \providecommand\BibTeX{{%
    \normalfont B\kern-0.5em{\scshape i\kern-0.25em b}\kern-0.8em\TeX}}}
\usepackage{tabularx}

\setcopyright{acmlicensed}
\acmJournal{PACMHCI}
\acmYear{2023} \acmVolume{7} \acmNumber{CSCW1} 
\acmArticle{125} \acmMonth{4} \acmPrice{}\acmDOI{10.1145/3579601}

\newcommand\anedit[1]{{\color{black} #1}} 

\begin{document}


\title[Deliberating With AI]{Deliberating with AI: Improving Decision-Making for the Future through Participatory AI Design and Stakeholder Deliberation}



\author{Angie Zhang}
\email{angie.zhang@austin.utexas.edu}
\affiliation{
  \institution{School of Information, The University of Texas at Austin}
  \country{USA}
}

\author{Olympia Walker}
\email{o.walker@utexas.edu}
\affiliation{
  \institution{Dept. of Computer Science, The University of Texas at Austin}
  \country{USA}
}
\author{Kaci Nguyen}
\affiliation{
  \institution{School of Business, The University of Texas at Austin}
  \country{USA}
}
 \email{kacinguyen@utexas.edu}

 \author{Jiajun Dai}
 \email{janetd@utexas.edu}
 \affiliation{
   \institution{School of Information, The University of Texas at Austin}
   \country{USA}
 }
 \author{Anqing Chen}
 \email{benjamin.c0427@gmail.com}
 \affiliation{
   \institution{Dept. of Electrical and Computer Engineering, The University of Texas at Austin}
   \country{USA}
 }
 \author{Min Kyung Lee}
 \email{minkyung.lee@austin.utexas.edu}
 \affiliation{
   \institution{School of Information, The University of Texas at Austin}
   \country{USA}
 }
\renewcommand{\shortauthors}{Zhang et al.}


\begin{abstract}
Research exploring how to support decision-making has often used machine learning to automate or assist human decisions. We take an alternative approach for improving decision-making, using machine learning to help stakeholders surface ways to improve and make fairer decision-making processes. We created "Deliberating with AI", a web tool that enables people to create and evaluate ML models in order to examine strengths and shortcomings of past decision-making and deliberate on how to improve future decisions. We apply this tool to a context of people selection, having stakeholders---decision makers (faculty) and decision subjects (students)---use the tool to improve graduate school admission decisions. Through our case study, we demonstrate how the stakeholders used the web tool to create ML models that they used as boundary objects to deliberate over organization decision-making practices. We share insights from our study to inform future research on stakeholder-centered participatory AI design and technology for organizational decision-making.
\end{abstract}


\maketitle

\section{Introduction} 
Past research has shown the effectiveness and advantages of using statistics and algorithms to make decisions---compared to humans, these systematic methods can use the same data to replicate prediction outcomes and in many cases produce similar or better accuracies \cite{dawes1989clinical, grove2000clinical, aubreville2020deep}. Drawing on this potential and in conjunction with advances in machine learning (ML) and artificial intelligence (AI), many systems are being developed to make more effective decisions at scale in both the public and private sectors---from predicting risk for homelessness \cite{toros2018prioritizing} and child maltreatment \cite{cuccaro2017risk} to managing work forces \cite{carey_smith_2016, silverman_waller_2015} and allocating resources such as donations or vaccines \cite{lee2019webuildai, matrajt2021vaccine}. Many of these systems either automate decision-making or present humans with recommendations at the time of decision-making. Despite \anedit{its promise of scalable decision-making}, AI and ML models can result in biased or unfair decisions and affect harm on communities. Addressing these issues, an active area of research investigates ways to create models to be less biased, fair, equitable, and more accountable to the community. These efforts include understanding how to design fair ML models \cite{kleinberg2018human, de2020case, mehrabi2021survey, van2019crowdsourcing}, algorithmic auditing to uncover harms \cite{wilson2021building, raji2020closing, buolamwini2018gender}, participatory and community-centered approaches for AI design \cite{lee2019webuildai, saxena2021framework, katell2020toward, shen2022model}, and constructing complementary relationships between AI and humans at the time of decision-making \cite{Jarrahi2018-bl, madras2017predict, keswani2021towards}. 

We propose an alternative approach aimed at improving human decision-making using ML \anedit{for selection or allocation decisions of organizations}. Rather than trying to improve the design or performance of \textit{algorithmic systems}, we explore creating and deliberating with ML models to improve and make fairer \textit{human decision-making}. Just as AI and ML have the potential to exacerbate unintended harms, human decision-making is not necessarily objective either \cite{cowgill2017algorithmic}, as evidenced by continued racial discrimination in hiring decisions and bail determinations \cite{bertrand2004emily, quillian2020evidence, arnold2018racial} and gender bias in evaluations or hiring \cite{macnell2015s, goldin2000orchestrating}. 

We propose that creating ML models with historical data can help reveal patterns of successes---a premise that many ML-decision systems are based on---but also missteps and weaknesses such as human or systemic biases, non-inclusive practices and classifications, and lack of diverse representation. An ML model can externalize these patterns by training on historical data to make predictions and display the results to people. Pairing models with reflection---to help people realize behaviors to sustain or change---and deliberation---to help people share perspectives and/or reach a common understanding---can enable people to generate and share ideas for improving human decision-making. \anedit{Reflecting and deliberating to reach this common ground is important for individuals of an organization to refine their shared organizational goals.}

As a first step toward this goal, we created a web tool, \anedit{Deliberating with AI}, in which stakeholders create and evaluate ML models so that they may examine strengths and shortcomings of past decision-making and deliberate over how to improve future decisions. The design of the tool draws from participatory AI design in order to support stakeholders in creating ML models, as well as research on deliberation and reflection to center introspection and discussion during the build and evaluation of ML models. The final outcome of this web tool is not intended to be a socio-technical system or human process that has been objectively \anedit{rid of biases}; instead, we frame our web tool as a method to help users deliberate how to improve future \anedit{personal and organizational} decision-making while also guiding them in participatory AI design. \anedit{We define organizational decision-making as decision-making for members of the same organization.} We apply this web tool to a context of people selection, having faculty and students use it to address admissions decisions. We report the interactions and discourse from user studies, describing how participants \anedit{used their ML models to deliberate and the ways they envision fair decision-making}. Specifically, the ML models helped them share their perspectives with one another and talk about the nuances of admissions decision-making as it relates to future decision-making practices.

Our work makes contributions to emerging literature on human-centered use of AI and organizational decision-making in the fields of computer-supported cooperative work and human-computer interaction. We first explain the design of our \anedit{Deliberating with AI} web tool to illustrate how it enables participatory AI design as well as ML-driven deliberation. We then describe our findings from applying our web tool to a specific use case, presenting how our participants used ML models as boundary objects to deliberate over organizational decision-making practices. Finally, we share insights from our case study to inform future research on stakeholder-centered participatory AI design and technology for organizational decision-making.

\section{Related Work}

To situate our approach, we first review prior work which uses AI and ML to automate or assist human decision-making. These studies focus on how to create fair and inclusive automated systems as opposed to centering fairness in human decision-making itself, and often do not contain elements of reflection and deliberation to encourage users to consider personal and systemic biases. For that reason, we next frame how reflection and deliberation can be integrated to advance fair human decision-making, identifying that while data-driven reflection has been used to help individuals assess personal data for health behavior insights, it has been less explored with individuals and groups for assessing data regarding others.

\subsection{Pursuing AI as a Complement to Human Decision-Making}
Efforts to support human decision-making with ML have often investigated how AI and humans can work complementary with one another, such as having ML models learn when to defer to humans \cite{madras2017predict, keswani2021towards} or designing AI analytic tools to augment human intuition \cite{Jarrahi2018-bl, calisto2021introduction, calisto2022breastscreening}. In some instances, researchers have tested if ML models can make better predictions than humans, even finding in some cases that accounting for additional human unpredictability, models can outperform human decision makers \cite{kleinberg2018human}. Others have focused on understanding the needs of \anedit{data scientists} and practitioners when creating fair AI systems in order to inform techniques that can help them \cite{holstein2019improving,madaio2022assessing} such as AI fairness checklists \cite{madaio2020co}, provenance for datasets \cite{gebru2021datasheets, hutchinson2021towards}, visualizations of model behaviors \cite{wexler2019if}, and toolkits for detecting and mitigating against ML unfairness \cite{bird2020fairlearn, bellamy2019ai, shen2021value}. Yet even with these tools, practitioners may still face challenges in how to use them in practice for specific use cases \cite{lee2021landscape}.

\subsection{Stakeholder Involvement in AI/ML design} 
One criticism of automated decision-making systems is that impacted stakeholders are rarely consulted in the design of them. The lack of involvement can lead to not only harmful systems \cite{angwin2016machine} but also lowered trust in individuals who perceive algorithmic unfairness \cite{woodruff2018qualitative}. In response, a line of research studies how to incorporate community members into the process, and whether community engagement or participatory approaches can lead to fair designs of ML. \citet{zhu2018value} and \citet{smith2020keeping} used Value Sensitive Algorithm Design to identify impacted stakeholders, explore their values, and incorporate their values and feedback to an algorithm prototype. Others have explored eliciting stakeholders' fairness notions to improve algorithms---\citet{srivastava2019mathematical} found that the simplest mathematical definition, demographic parity, aligned best with participants' fairness attitudes while \citet{van2019crowdsourcing} found that having diverse groups judge the fairness of model indicators can lead to more widely accepted algorithms. Researchers have also explored how to design AI or ML models with stakeholders, \anedit{such \citet{lee2019webuildai}'s participatory AI framework to create ML models and \citet{cheng2021soliciting}'s stakeholder-centered framework that probes their participants' fairness notions to design fair ML.} Others such as \citet{holstein2019co} and \citet{zhang2022algorithmic} have used co-design in order to work alongside impacted stakeholders to create AI or AI interventions.

While participatory methods aim to make AI/ML design more inclusive and equitable, designing responsible ML decision-making tools or processes for stakeholders not trained or versed in ML ("non-experts") is a uniquely challenging task. In order to assist stakeholders in understanding automated decisions or creating ML models, researchers and organizations have \anedit{created} various tools and interfaces. \citet{yang2018grounding} engaged with non-experts \anedit{who used} ML to surface design implications of ML, while \citet{yu2020keeping} and \citet{ye2021wikipedia} created data visualizations to convey algorithmic trade-offs in more understandable ways for designers and other users. Similarly, \citet{shen2020designing} tested different representations of confusion matrices to support non-experts in evaluating ML models. These studies aim to lower the barriers for non-experts to participate in and advance fair algorithmic decision-making. Ultimately though, their objective is the creation of automated decision-making systems. We draw inspiration from these works, but we focus on improving \textit{human decision-making} processes as our outcome instead, through the use of machine learning. Our design is further motivated by literature on reflection and deliberation which we explain next.

\subsection{The Role of Reflection and Deliberation}
Self-reflection is a process to support individuals in making realizations about themselves and even changes to their behavior. Researchers have explored how to aid self-reflection through design of technologies and strategies that help individuals collect and probe their own health data \cite{li2011understanding, lee2015personalization, choe2017understanding}. \citet{lee2015personalization} observed how a reflective strategy (e.g., reflective questions) increased participants' motivation to set higher goals. \citet{choe2017understanding} found that data visualization supports helped participants recall past behaviors and generate new questions about their behaviors to explore in the data. Similarly, in our tool we include designs to support self-reflection such as reflective questions and data visualizations, drawing on Sch\"{o}n's Reflective Model, so participants can engage in reflection-in-action and reflection-on-action while creating their ML model \cite{schon2017reflective}. This is intended so that they may critically analyze their decision-making preferences and historical data for insights. But in contrast to prior work, the data presented to participants is not their personal data and the insights we ask them to draw are not related to their own health behaviors: instead participants review aggregated, anonymized data to reflect on fairness and bias in relation to organizational decision-making.

While reflection allows individuals to contemplate their own experiences and decision-making, it does not necessarily take into account group-level insights and preferences for decision-making. For this, we turn to deliberation to encourage input and participation from all members. Group deliberation can be traced back to public deliberation or deliberative democracy, where citizens gather to discuss policies that will impact them \cite{fishkin2002deliberative}. More recent studies exploring online deliberation demonstrate how it can help increase the accuracy of crowdworking tasks \cite{schaekermann2018resolvable,drapeau2016microtalk,chen2019cicero}, improve perceptions of procedural justice \cite{fan2020digital}, and support \anedit{consensus building} amongst participants \cite{xie2020chexplain,schaekermann2019understanding,lee2020solutionchat,van2019crowdsourcing}.

Scholars suggest the importance of combining deliberation with reflection, such as \citet{ercan2019public} who describe how accompanying deliberation with reflection is needed to support intentional, deeper conversations and \citet{goodin2003does} who describe how political deliberation consists of not just formal public deliberations but also internal reflections and informal deliberations. With this in mind, we constructed the \anedit{Deliberating with AI} web tool to incorporate both. Past research \anedit{has often integrated} a user's reflection with asynchronous public deliberation by having users read the stances of others \cite{kriplean2012supporting, schaekermann2018resolvable, fan2020digital, van2019crowdsourcing} although \citet{schaekermann2019understanding} incorporated face-to-face and video modes as well. In our approach, we iterate between synchronous deliberation and reflection such that deliberation can allow users to hear each others' perspectives to reflect over and reflection can help them make new realizations to share with the collective.

\section{Participatory AI Design and Stakeholder Deliberation for Decision-Making with AI}

\anedit{We first discuss the goals of and design choices that went into Deliberating with AI, explaining our approach of participatory AI design to center stakeholders in the design of ML models. Reflection and deliberation allow stakeholders to surface normative values while creating ML models, and ML models themselves augment stakeholder deliberation and address decision-making practices.} We then explain the implementation of these design principles in the web tool.

\subsection{Goals for Using the Deliberating with AI Web Tool}
\anedit{Our objective for the Deliberating with AI web tool is supporting organizations in making decisions that are fair, inclusive, and effective for present and future communities. By this, we mean outcomes or decisions that impact different communities similarly, do not discriminate based on sensitive attributes such as ethnicity, and support the organization's goals. We also intend for users of the tool to define for themselves what they believe the organization's ideal outcomes should be. We drew inspiration for our tool design from \citet{van2021machine}'s ethnographic study about data scientists building a hiring ML system for an organization, where after contextualizing the meaning of hiring data with HR staff, the organization and data scientists found historical hiring decisions potentially reflected anchoring bias.}

\subsection{Embedding Deliberation throughout the AI/ML Design Process}

Prior work has investigated how to assist stakeholders or practitioners at specific steps of an ML model building pipeline, such as model evaluation \cite{ye2021wikipedia, wexler2019if, cheng2019explaining, shen2022model}. However, little work has covered engaging stakeholders in the whole ML pipeline, from creation to evaluation. We explore this in the design of a web tool such that users create and evaluate models by following four standard stages of an ML building pipeline: data exploration, feature selection, model training, and model evaluation. 

We draw inspiration for our tool from participatory AI design \cite{lee2019webuildai, ye2021wikipedia, shen2022model} due to its premise of centering community members or impacted stakeholders in the process of AI design. \citet{lee2019webuildai}'s study found that as a result of creating individual ML models, participants gained awareness of gaps in their organization's decision-making. Likewise, we are interested in whether having users create and evaluate ML models can help them surface patterns of organizational decision-making (e.g., gaps, shortcomings, successes) to inform future practices. 

Our \anedit{Deliberating with AI} web tool is intended to be used by an organization that has to reach consensus on criteria for a selection or allocation problem and has data about past decisions. Often though, individual preferences are not homogeneous within an organization, so the organization needs a way to elicit and balance the preferences for a group. To address individuals' conflicting preferences and the potential of historical data to reveal past decision-making patterns, our web tool facilitates a participatory process so that members of an organization can build and evaluate ML models to \anedit{then review and discuss} what they envision fair decision-making to be. \anedit{We choose to use ML models for identifying biases and potential harms of past decisions as they can help participants identify what factors played a role in past human decisions in a more cognitively digestible format compared to looking at historical data on its own.}

\anedit{In this section, we elaborate on how incorporating reflection and deliberation in our tool can support surfacing normative values at each ML model building stage and how users' ML models can help them deliberate over decision-making practices.}

\subsubsection{Using Reflection and Deliberation to Surface Normative Values When Creating ML Models}

Although building an ML model is traditionally seen as a very technical task and done primarily by \anedit{data scientists} with limited or no stakeholder input, an ML model built this way risks biases and inflicting disparate harms on populations \cite{angwin2016machine, martinez2021secret}. To counter that, researchers have argued for a socio-technical approach in constructing AI/ML that incorporates stakeholders into the process to inform its functions \cite{smith2020keeping, d2020data, aragon2022human}.

In line with this, the design of our \anedit{Deliberating with AI} web tool is influenced by literature on public deliberation which emphasizes participation and exposure to diverse perspectives \cite{fishkin2002deliberative}, as well as methods for and models of reflection which center introspection of one's actions and surroundings \cite{lee2015personalization, choe2017understanding, schon2017reflective}. Based on literature supporting the use of deliberation and reflection together to deepen deliberation \cite{ercan2019public, goodin2003does}, our tool iterates between synchronous deliberation and reflection to allow the two modes to work together and enhance one another.

Participants begin the model creation process through data exploration. Typically data exploration is used by \anedit{data scientists} to evaluate things such as the distribution of data, patterns that indicate what algorithm to use for model training, and resolve missing values. However, while \anedit{data scientists} may be well versed in choosing suitable training algorithms for the dataset, stakeholders with lived experience or domain knowledge can be engaged to explore the data for its practical meaning---e.g., signaling variables that \anedit{data scientists} may otherwise overlook, providing guidance on the meaning of missing data. 

In feature selection, reflection and deliberation play integral parts for creating a model that reflects stakeholder values. When \anedit{data scientists} select features for a model without input from stakeholders with domain expertise or lived experiences, they risk creating a model that is unrepresentative and unsupportive of stakeholders who will be impacted \cite{grgic2018beyond, lee2019procedural}. \anedit{These principles can draw out contexts not obvious to data scientists such as specific reasons for support or concern of using a feature, how features are related to one another, and situations or exceptions that may change how a feature should be treated.}

Model training requires very technical knowledge from \anedit{data scientists} who must match and test different ML algorithms to optimize the model's performance. However, there are still normative behaviors of models that deliberation with stakeholders can benefit. For example, conveying transparency and explainability of different models to stakeholders remains a challenge for researchers. Engaging in reflection and deliberation at this stage may help researchers clarify the specific questions stakeholders have that researchers can address to improve model transparency or explainability. 

Finally, model evaluation is traditionally done by evaluating performance metrics such as accuracy, precision, and recall. Having stakeholders reflect and deliberate at this stage can complement standard ML metrics \anedit{with the expectations and values stakeholders use in practice to assess a model's performance}. Stakeholders can also reflect on alternate ways they wish to review models. Model evaluation has been the site of extensive research efforts to incorporate stakeholder feedback in AI design \cite{wexler2019if, ye2021wikipedia, yu2020keeping, shen2022model}. We distinguish our study from prior work by emphasizing that our web tool is specifically intended to support stakeholder deliberation, a goal not emphasized in past work. We also incorporate a new aid to make model evaluation more accessible for users that contrasts from the visualization aids of \cite{ye2021wikipedia} and \cite{wexler2019if}---our web tool's Personas screen allows users to retrieve anonymized profiles from the dataset and review the model's decision, serving as an alternate method for users to assess the model's performance at the individual level.

\subsubsection{Constructing Individual and Group Models to Define Decision-Making Practices and Augment Deliberation} 

When using \anedit{Deliberating with AI}, users create two types of models: an individual model on their own and a group model with others. They begin with individual ML models which allow them to actualize their abstract beliefs outside of other users' influence. Then they work collectively, sharing which features they want to include and why, in order to create a group model that represents decision-making informed by all group members. Models are not only useful for helping users make concrete the ideas they have for decision-making, they can also act as boundary objects \cite{star1989structure}, or "common frames of reference" \cite{cai2021onboarding} for users to structure their discussions. Structured discussion has been found to make deliberation more effective by guiding users to stay on task \cite{farnham2000structured}, increasing the quality of arguments users give to support their beliefs \cite{drapeau2016microtalk, schaekermann2018resolvable}, helping users recognize how disagreements arise \cite{schaekermann2019understanding}, and assisting users in keeping track of points made in text-based discussions \cite{lee2020solutionchat}. To capitalize on these potential benefits of structured discussion, in our tool, ML models as boundary objects can assist users in structuring group discussions so they can gain the most from deliberation to formulate ideas for decision-making practices. 

\subsection{Web Tool Implementation}
\label{webtool}
\begin{figure}
\includegraphics[width=.9\textwidth]{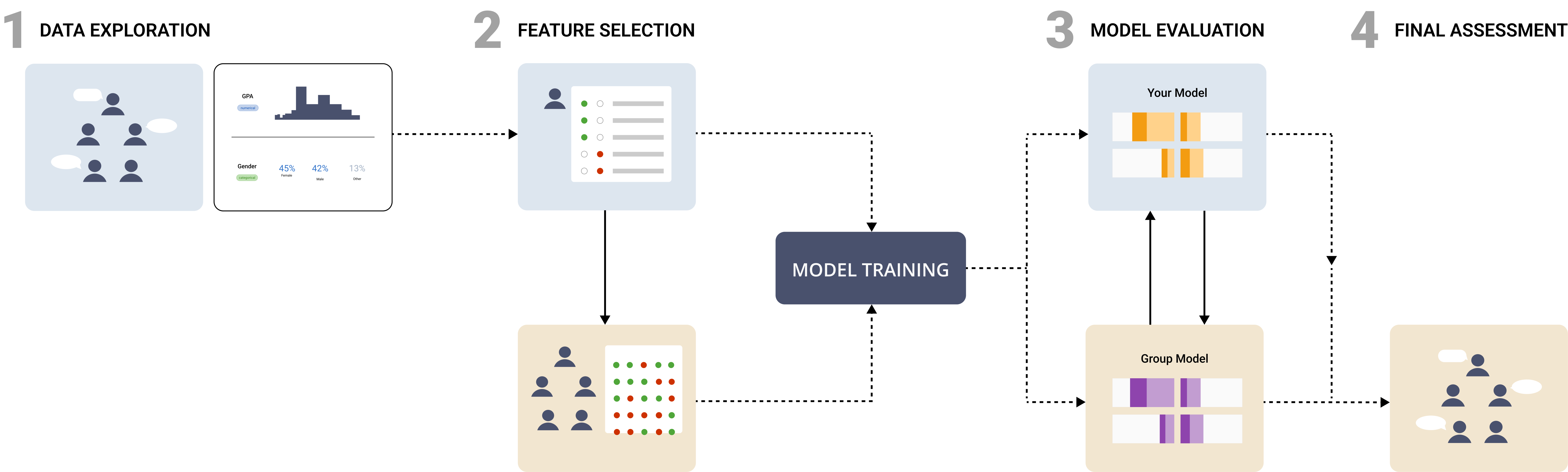}
\caption{\anedit{Deliberating with AI Session Flow. 1.) The session is facilitated primarily via the web tool and begins with Data Exploration where as a group and then individually, participants review the data. 2.) Next on the web tool, participants make feature selections to be applied to their individual models before moving to a MIRO board to deliberate feature selections as a group. These feature selections are used in training unique individual models and one group model. 3) The participants evaluate their model using the metrics and visuals provided by the web tool, both individually, and then 4) discussing as a group before ending with an exit interview.}}
\label{fig:Flow}
\end{figure}

We designed the \anedit{Deliberating with AI} web tool according to the principles above. The web tool begins with an overview to explain AI and ML, and introduces the problem domain. We emphasize to users that the goal is not to create a perfect model for automated decision-making, but to identify gaps and patterns of historical decision-making and imagine alternate ways to leverage ML to improve future decisions. (See Fig. \ref{fig:Flow} for the web tool and session flow.)

\subsubsection{Data Exploration} The web tool first allows users to collectively examine past data (i.e., input and decision outcomes) and shows a predictive model trained on the data ("All-Features" Model) which displays factors from the data that played a role in outcomes and to what degree. This design is to help users reflect and share thoughts on past decision patterns. To assist introspection, the web tool asks users questions to prompt self-reflection around the problem space. To encourage users to explore and become familiar with the dataset, but conscious of how overwhelming data-related activities can be, we designed the tool to display factors in a digestible way (see Fig. \ref{fig:dataexplore}): the interface combines explanations side by side with activities, with tabs to separate the activities in each section. In data exploration, the web tool displays each factor using its name, a visual to give a quick indication of its distribution, and summary statistics about its distribution.

\subsubsection{Feature Selection for Building Individual Models}
\begin{figure}[h]
\includegraphics[width=\textwidth]{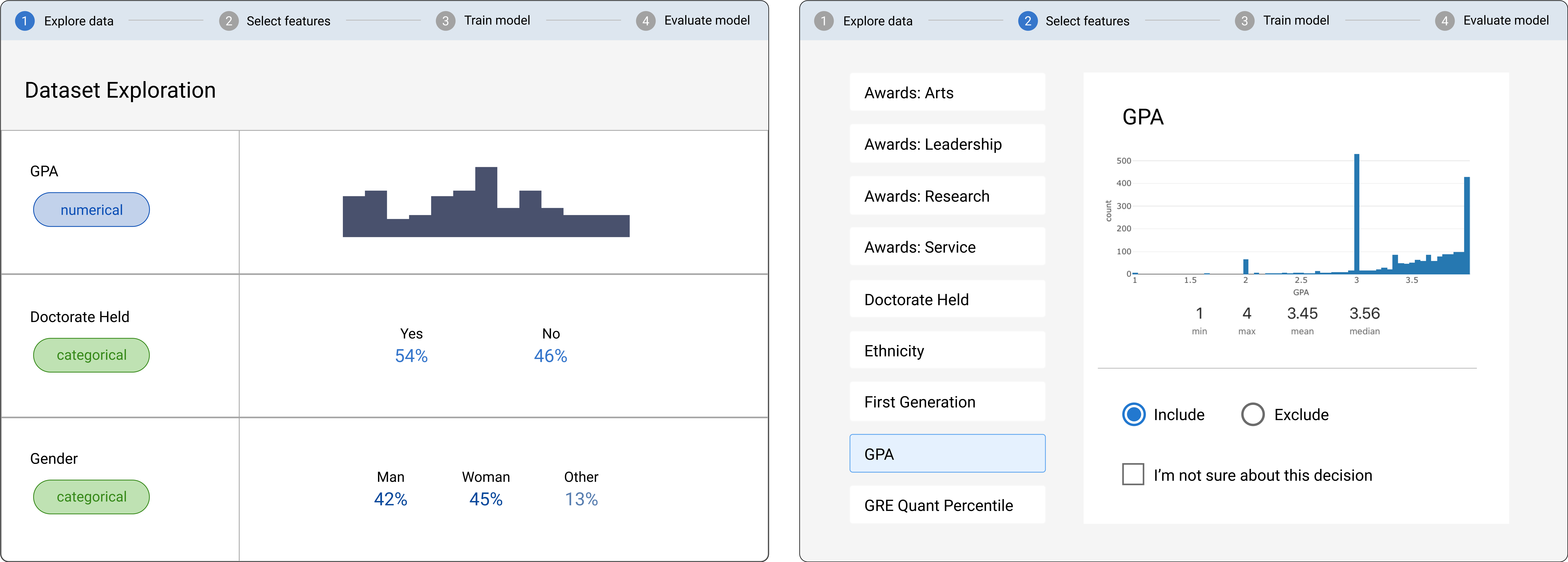}
\caption{\anedit{Screens of Deliberating with AI web tool. Left: Data Exploration screen that participants use to review each feature and its distribution of values. Right: Feature Selection screen that participants use to decide whether to include each feature and why.}}
\label{fig:dataexplore}
\end{figure}

Following the initial exploration of data, users select what factors, or features, they believe should be used, regardless of the extent to which those features played a role in the "All-Features" Model \anedit{(Fig. \ref{fig:dataexplore})}. This design helps users externalize their ideal decision criteria by having them choose to include or exclude each feature, share their reasoning, and/or flag if they are unsure. \anedit{We ask these in the web tool to encourage} self-reflection and document details to discuss during group deliberation later. Users can also explore the dataset by reviewing bivariate distributions between pairs of features in order to generate hypotheses to explore later. For example, in our case study, a user can select the features \textit{Ethnicity} and \textit{GPA} to view box plots of GPA by ethnicity.

\subsubsection{Feature Selection and Deliberation for Building a Group Model}
\begin{figure}[H]
\centering
\begin{minipage}{.48\linewidth}
    \includegraphics[width=\linewidth]{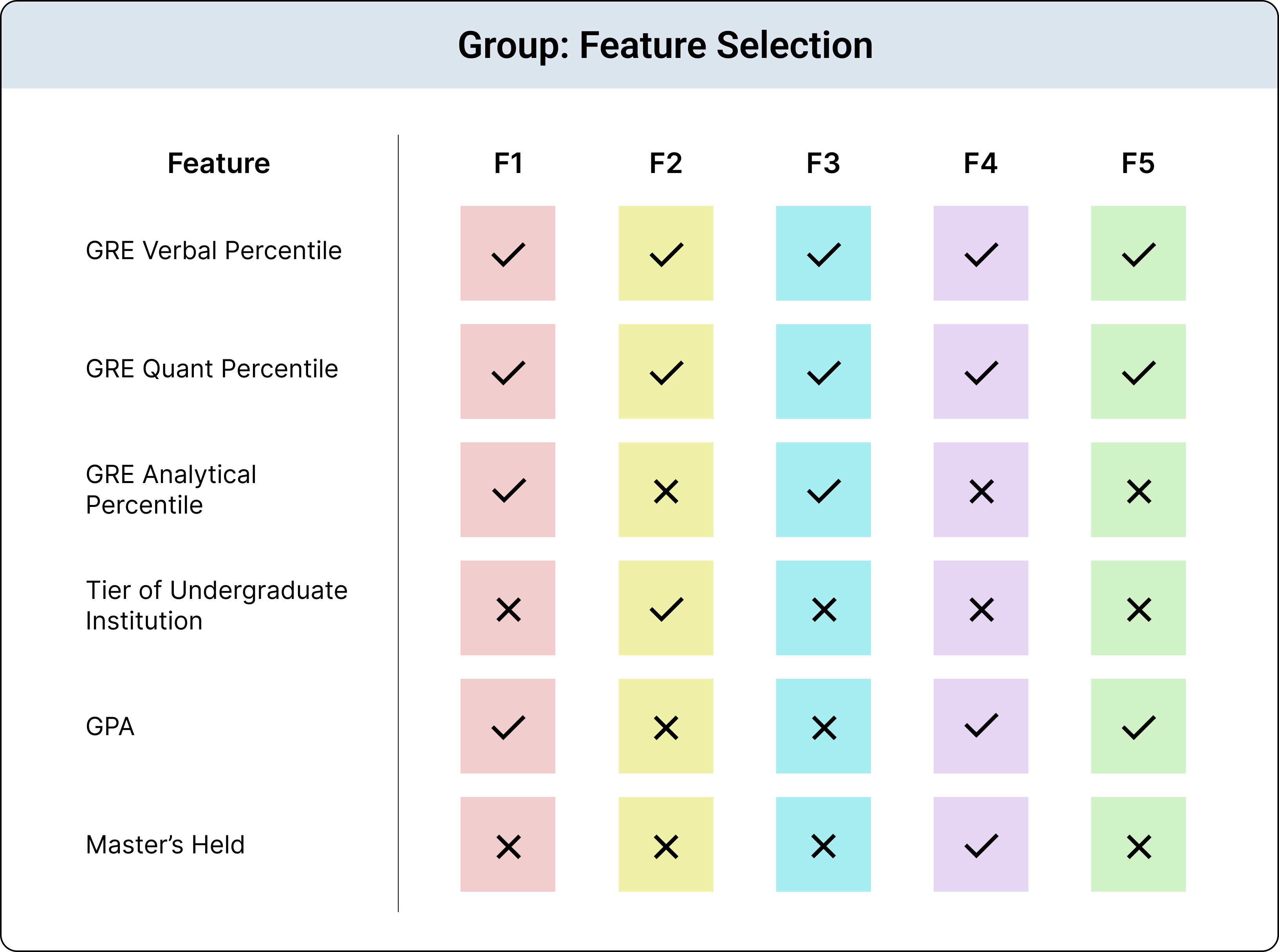}
    \caption{\anedit{Layout for deliberation session of Feature Selection in MIRO.}}
    \label{MIRO1}
\end{minipage}
\hfill
\begin{minipage}{.48\linewidth}
    \includegraphics[width=\linewidth]{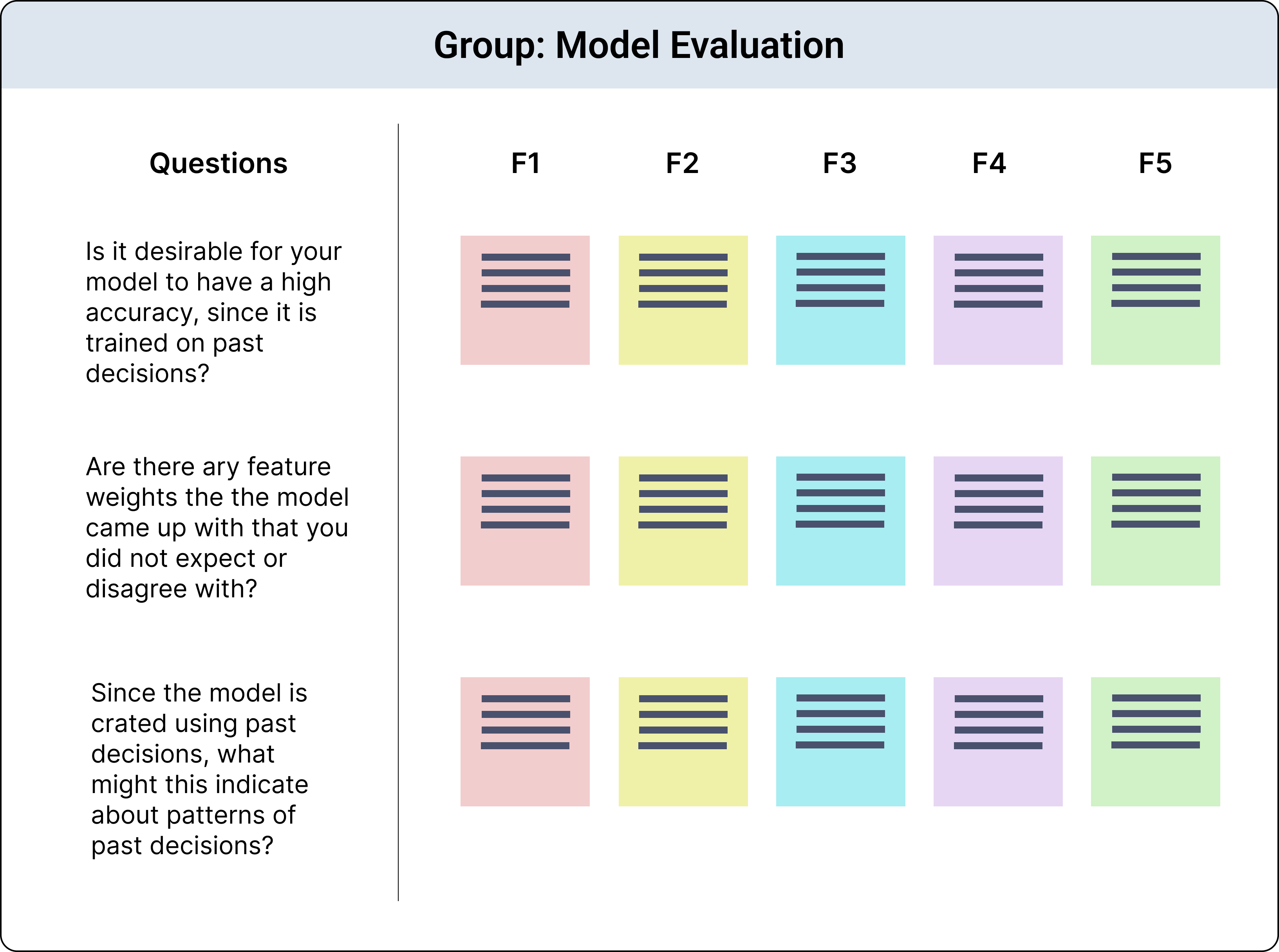}
    \caption{\anedit{Layout for deliberation session of Model Evaluation in MIRO.}}
    \label{MIRO2}
\end{minipage}
\end{figure}

To support deliberation and group model construction, the web tool compiles the data from the previous step, feature selection for building the individual model, into a flat file: for each feature, the file displays each user's decision (include or exclude), their reasons, and whether they were unsure. This file can be imported into an online whiteboard tool such as a MIRO\footnote{https://miro.com/} board to display the results to the group of stakeholders and provide a basis for deliberation. \anedit{Users deliberate to finalize a set of features for the group model \anedit{(Fig. \ref{MIRO1})}.} The rules for establishing consensus per feature is not embedded in the tool, thus this can be a decision determined by the facilitator and/or participants depending on group size, opinions, and time constraints.

\subsubsection{Model Training}
Once consensus on features for a group model has been finalized, the results are input into the tool by the facilitator. Participants return to the web tool for the remaining \anedit{Deliberating with AI} process. On the administrator interface of the tool, the facilitator trains the group and individual models. 

The features selected by users are used to train their unique individual models; \anedit{the group model is trained using the features selected by the group through deliberation.} While models are trained, users watch a video to get an understanding of a basic ML model training process. We simplify the details while retaining the core steps to balance not overwhelming the user with complex concepts while providing sufficient information of how a model is trained. 

One design consideration we faced was the trade-off of explaining ML model/classifier types for users to deliberate over vs. focusing on one straightforward model to introduce users to ML concepts. Although we ultimately implemented the latter thus users do not engage in model selection, the potential impacts of including reflection and deliberation for model selection are an important consideration, especially given different models can make different errors \cite{narkar2021model}, which would impact the perceptions and ideas of users engaging with the \anedit{Deliberating with AI} web tool.

\begin{figure}[h]
\includegraphics[width=\textwidth]{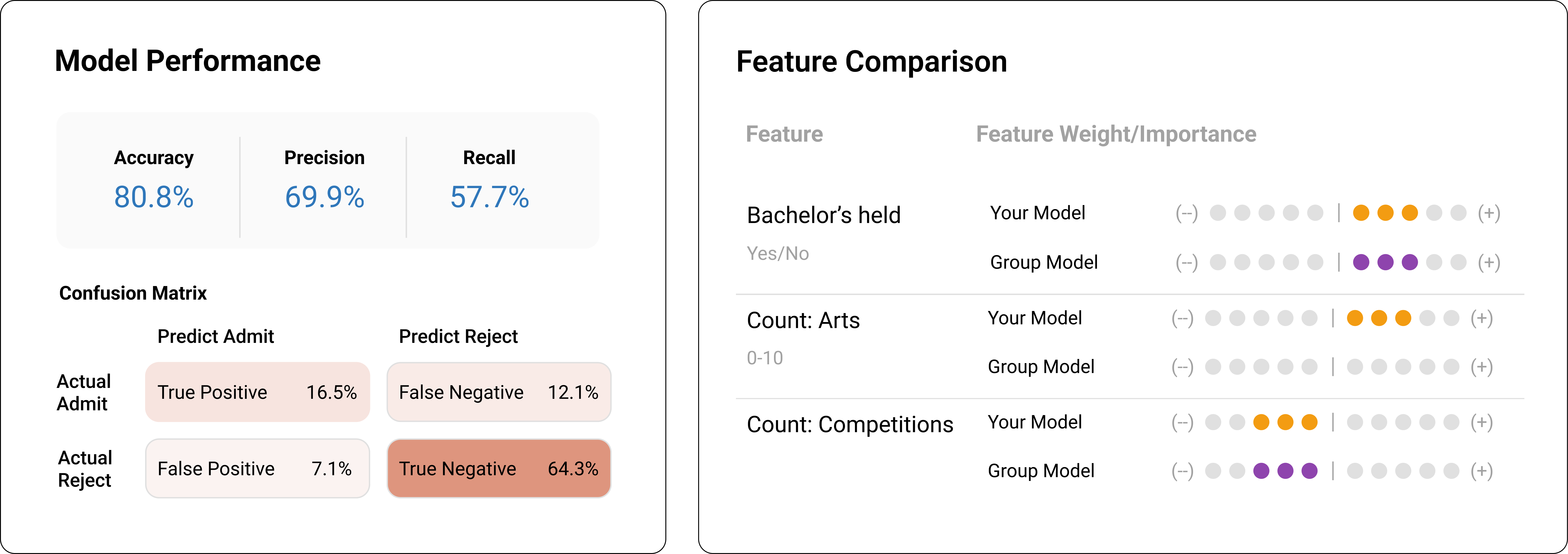}

\caption{Left: Model Performance screen explains traditional ML metrics of participant models. Right: Feature Weight Comparison screen shows weights learned for each feature in ML models.}
\label{fig:performftweight}
\end{figure}

\begin{figure}[h]
\includegraphics[width=\textwidth]{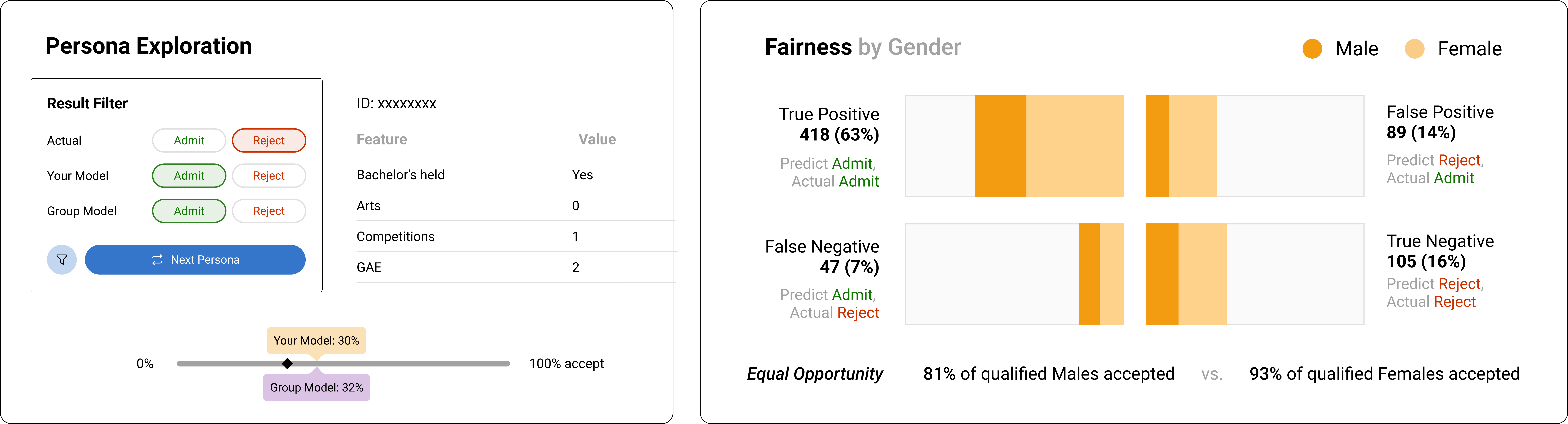}
\caption{Left: Personas screen shows ML model predictions compared to real historical decisions on anonymized applicants. Right: Fairness screen shows ML model compliance with specific fairness definitions.}
\label{fig:personasfairness}
\end{figure}

\subsubsection{Evaluating Individual and Group Models}
When evaluating models, all users assess two models: 1) the same group model and 2) their unique individual model. The web tool provides users with multiple ways of evaluating the models so that they can understand what factors ML models use, kinds of errors ML models can make, and how fairness can be conceptualized in ML. This is to provide people with an understanding of ML capability, risk, and associated trade-offs, so that they can imagine ways to leverage ML to strengthen future decision-making if they desire. On each evaluation screen (explained below), a reflective question prompts users to think about the costs and benefits of human vs. automated decision-making. For example, on Personas, the question reads, "If the prediction from the models and the actual admission decision differs, which do you agree with and why?" On each screen, group deliberation allows participants to share responses to the reflective question and reactions to the model and tool activities.

\textit{Feature Weights.} The web tool displays the feature weights of the individual vs. group model. If a user did not select a feature that the group did or vice versa, no weight will be shown for the corresponding model (Fig. \ref{fig:performftweight}). 

\textit{Personas.} To provide a tangible idea of who is receiving "correct" (model results align with historical decisions) vs "erroneous" (model results do not align with past decisions) predictions, this screen \anedit{(Fig. \ref{fig:personasfairness})} allows people to retrieve personas---profiles of anonymized individuals displaying their values for all features. Users can filter personas based on admittance or rejection by their ML models and/or past decision makers. Matching profiles are displayed with a score and model confidence, and additional filtering options allow users to narrow down the results by features such as gender, ethnicity, etc., with up to two features at a time. This enables users to probe how specific groups of applicants may have been impacted by the model and/or past decision makers.

\textit{Model Performance.} Users can see metrics (e.g., accuracy, recall) for their individual vs. group model \anedit{(Fig. \ref{fig:performftweight})}. We include a contextualized confusion matrix similar to \cite{shen2020designing} to help users understand terms like “False Positives” in the context of a specific problem domain, augmented with textual explanations inspired by \cite{yu2020keeping}. We also apply visual changes to the display so if the user selects to see how a metric is calculated, only the relevant quadrants of the confusion matrix remain on the screen.

\textit{Fairness.} We provide two definitions of mathematical fairness---equal opportunity and demographic parity---as an introduction for users to consider what fair or unfair model outcomes look like in ML. We include graphs based on confusion matrix visualizations of \cite{yu2020keeping} to demonstrate whether disparities exist in group treatment under different fairness definitions \anedit{(Fig. \ref{fig:personasfairness})}.

\subsubsection{Technical Implementation Details}
The web tool is built with React and Material-UI on the frontend, and Flask and MongoDB in the backend. The plotly.js package is used for displaying graphs to users. In addition to the user functions of the main web tool, an “Admin View” provides functions for facilitating the session (e.g., downloading a flat file of all user feature selections), abstracting the needs of facilitators from highly technical tasks such as dropping a database or creating a JSON file from a JavaScript object. For group deliberation, we use MIRO boards to take advantage of existing technologies designed for virtual collaboration. 

While it can be used by an individual user, this tool is designed to be used by multiple people who are in a same decision maker role, or who are affected by the decisions, so that the resulting recommendations reflect multiple people's perspectives and to mitigate the potential effects of imbalanced power dynamics and users suppressing their opinions amongst mixed stakeholder types.

\section{Case Study: University Admissions Review Process}

We applied \anedit{Deliberating with AI} to the context of the master's admissions review process at a public university in the United States. Universities are increasingly exploring options to use AI in recruiting and reviewing applicants \cite{kelliher_2021, newton_2021}, making this a pressing domain for us to turn our attention to.

\subsection{Study Context}
\subsubsection{The Challenges in the Master's Admissions Review Process}

The master's admissions review process in the United States is an inexact, ambiguous, and sometimes contentious undertaking \cite{v2014539, kramer2019timeline}. Today, most schools use holistic review where admittance is based on the entirety of an individual’s application (e.g., academic performance, essays, individual attributes) rather than a single criteria. Intended to improve fairness of the admissions review process, in reality, holistic review can be opaque and inconsistent---perhaps unsurprising as it resulted from a systematic effort by elite colleges to exclude top-scoring Jewish students \cite{karabel2005chosen}.

Master’s review committees face numerous challenges: 1) holistic review means different things to different people contributing to the inconsistency of decision-making \cite{bastedo2018we, kent2016holistic}, 2) members often lack guidance on how to properly weight and assess a multitude of criteria such as GRE scores and GPAs from different institutions \cite{diminnie1992essential}, 3) the holistic review process is very demanding on reviewers' time \cite{talkad2018making}, a growing concern as the volume of graduate applications has continued to rise in the United States, and 4) reviewers can and often do bring unconscious biases when assessing applicants which may exacerbate gatekeeping tendencies in admissions \cite{posselt2016inside, talkad2018making}. 

In the past, to address some of these issues, people have created automated tools for assessing applicants to enable consistency and ease the burden on human reviewers \cite{waters2014grade, pangburn_2019} \anedit{and even for applicants to predict their potential for admittance to specific graduate schools \cite{acharya2019comparison}}. This increased use/interest in predictive analytics has understandably raised the concerns of many due to the potential of AI to inflict disparate harms on populations \cite{martinez2021secret, brown2019toward}. However, even without automated tools, the master’s admissions review process is still inherently saddled by potential reviewer bias in the assessment of applicants. 

For these reasons, we turn our attention to a case study exploring how to improve human decision-making in master's admissions. Limited research explores computer-supported cooperative work around this domain, with exceptions of \cite{sukumar2018visualization} and \cite{metoyer2020supporting} exploring the ways that visualizations can support the work of review committees---in contrast, we focus on how ML models can support the deliberation of stakeholders to improve decision-making practices. Importantly, the final outcome of this web tool is not intended to be a socio-technical system or human process that has been objectively rid of biases. Instead, we frame our web tool as a preliminary method for assisting humans in identifying biases and potential harms in historical data in order to determine how decisions in the future should be made.

\subsubsection{Admission Dataset}
We apply this web tool to the context of the master’s admissions review process of a specific discipline at a public university. We submitted an IRB that allowed us access to historical admissions data, and we worked with the graduate school office to obtain anonymized applicant data with admission decisions. We were unable to obtain letters of recommendation or essays due to privacy reasons and thus could not include them. 

We describe the steps that were necessary for us to be able to apply \anedit{Deliberating with AI} to our use case to provide more details around our study. These steps also illustrate points of consideration for future use cases to keep in mind when using \anedit{Deliberating with AI}. Namely, we had to make key decisions around 1) data pre-processing in order to curate a usable dataset for the tool to ingest, 2) the procedure for reaching consensus during group deliberation, and 3) model type selection.

\textit{Data pre-processing.} We removed incomplete or pending applications, anyone who was not a master’s applicant, and applicants with GRE scores before scoring changes in 2011. The final dataset consisted of 2207 applicants, from Fall 2013 to Fall 2019 cycles. Next, we removed irrelevant columns (e.g., timestamps) and engineered features such as using parental education to construct \textit{First Generation}, and having researchers code features using external knowledge, such as \textit{Tier of Undergrad Inst.} (See full details on Table \ref{table:TableOfFeatureLogic2} in the Appendix.) Although gender and ethnicity are not permissible for use in admissions decisions, we included them for deliberation purposes. This resulted in the 18 features shown in Table \ref{table:TableOfFeatures2}. \footnote{One way to measure the tool's effects on human decision-making is through performance measures to evaluate pre- and post-tool decisions. We explored various performance measures---e.g., whether a student had a job offer upon graduation, what enrolled students' final graduate GPAs were. However, we faced challenges in obtaining complete performance measure data---e.g., job details for graduating students is not collected, post-application data is not collected about applicants who were rejected or admitted but did not enroll. While this is a limitation of our specific case and dataset, the integration of performance measures for future use cases can enable a more measurable impact of this tool.}

\textit{Procedure for reaching consensus in group deliberation.}
The ideal scenario for group deliberation without constraints is to discuss the inclusion/exclusion of each feature until unanimous consensus is reached. However, unanimity is not always feasible even with deliberation. Due to timing of our sessions, the decision to include a feature was based on the majority of votes once discussion abated, or if no noticeable agreement was observed from ongoing conversations. Although not ideal, the primary facilitator acted as a tiebreaker to ensure enough time for the remaining activities. Upon reflection, the authors recognize alternatives that could have been used, e.g., having participants vote on their preferred method for consensus, although this would have required longer time commitments from participants.

\textit{Selection of machine learning model.} We trained different models with our dataset (i.e., decision tree, ridge regression, lasso regression, and linear regression), achieving a similar baseline accuracy of 75\%-80\% for all models when using all 18 features. We ultimately chose to use linear regression with a 70/30 train-test split because it had the highest accuracy of our models, is fast to train, provides us with easy-to-understand insight into the model through feature coefficients, and is comparatively easier to explain to participants than more complex models.\footnote{Although a decision tree is an explainable ML model, in practice, it can quickly become time-consuming and complex with its many branches.} The fast training time of linear regression allowed us to create models during sessions in real time, and the feature coefficients and straightforward nature of the model made it a preferable choice to use with participants.

\subsection{Participants: Decision Makers and Decision Subjects}

In order to explore how \anedit{Deliberating with AI} can affect fair and responsible human decision-making, we conducted group sessions with nine student participants (decision subjects) and seven faculty participants (decision makers) at a public university in the U.S. Historically, master’s admissions review committees have consisted solely of faculty members. Via email, we recruited those with review committee experience and those without for varying perspectives. Although students do not currently serve on the review committee, we felt it was important to include the perspectives of impacted stakeholders. We recruited master’s students by posting a sign-up with details to a general Discord channel and Pride Discord channel for current students. We held separate sessions for faculty and students so as to not exacerbate the power differentials between the groups, especially given the possibility that some students may have taken or may take a faculty participant’s course in the future.

To protect the identities of the participants, we only provide aggregate statistics. Participants’ ages ranged from 18-74, and 62.5\% identified as female (37.5\% as male). Three identified as Hispanic, Latino, or Spanish origin. Ten described their ethnicity as white, three as Asian, one as American Indian or Alaska Native, one as Black or African American, and one answered "Other-Jewish Tejana, mixed race”. Nine participants (56.25\%) reported a Bachelor’s degree as their highest degree completed at the time of the study, one (6.25\%) reported a Master's degree, and six (37.5\%) reported a Doctorate. We also asked the participants about their current knowledge on programming and computational algorithms \cite{lee2017algorithmic}. The average participant programming knowledge reported was 2.5 (between 2-"A little knowledge-I know basic concepts in programming" and 3-"Some knowledge-I have coded a few programs before": SD = .89). The average participant computational algorithm knowledge was 2.1 (between 2-"A little knowledge-I know basic concepts in algorithms" and 3-"Some knowledge-I have used algorithms before": SD = .62).

\subsection{Procedure}
We conducted four virtual Zoom sessions: two with faculty members and two with students. The participants interacted with the tool to create and evaluate admissions decision-making models. Sessions lasted 2-2.5 hours each, with 3-5 participants each, and participants were compensated with \$80 Amazon gift cards for their time. Group segments took place in the main room; individual segments were held via breakout rooms where each participant worked on the tool with a facilitator who prompted think-aloud or interview questions so that participants could share their reflections and describe experiences they may not have wished to share in a group setting.

The facilitator began with an overview of the activity while participants opened the web tool on their computers. In the sessions\footnote{The first two sessions began with Data Exploration. Upon reflection from session debriefs, the facilitator modified the remaining two sessions so that displaying the "All-Features" Model preceded this.}, the facilitator displayed the feature weights of the "All-Features" Model, a linear regression model trained with all 18 features of the dataset to familiarize participants with working with ML models. Participants then walked through Data Exploration. Next, participants worked on Feature Selection for their individual model in breakout rooms. Everyone returned to the main room to discuss feature selections for the group model, \anedit{deliberating over how specific features impact outcome fairness}. The facilitator exported the participant feature selections into a shared MIRO board (see Fig. \ref{MIRO1}) and guided participants through deliberation. 

Once completed, the group's feature selections were input into the tool, and all models were trained. Everyone watched the Model Training video together before moving onto Model Evaluation as a group. Participants also explored the Personas screen of Model Evaluation individually with their facilitator guiding them through how to use the screen so as to test it without distractions. Everyone returned in the main room to share the personas they explored or patterns they observed. The session concluded with wrap-up interviews in breakout rooms and an exit survey on Qualtrics. 

To ensure participants understood the tool as they used it, facilitators asked questions in breakout rooms and in the main room to solicit verbal confirmation or questions from participants. For example, to ensure participants understood how the model actually worked, the facilitator guided them through the Feature Weights screen of the Model Evaluation stage. Participants showed understanding of how the model worked through their interpretation of the weights, whereby positive larger weights indicated features that had a stronger impact on the model's acceptance of an applicant.

\subsection{Analysis} All sessions were screen-recorded using Zoom. Recordings were transcribed in Otter.ai with errors fixed manually. Using the qualitative data analysis method \cite{patton1990qualitative}, the primary researcher reviewed transcripts and MIRO board text, and generated initial codes based on the tool components and questions asked during sessions; and the codes were then discussed during meetings with other researchers and further synthesized.

\section{Findings}
Our participants were thoughtful in their reflections and discussions, indicating genuine interest in the subject matter and explorations of the tool. \anedit{Below, we describe the impact of the tool on improving decision-making through participant interactions with the web tool components and session activities. This included not only the use of ML models to share perspectives around admissions and participant suggestions for new features to improve fairness and inclusion, but also alternate ideas for how to improve admissions decision-making.} Faculty participants are denoted with an F-prefix, and students with a P.

\subsection{Identifying Prospective Applicants}

We asked participants to both reflect on and discuss as a group what matters to them in applicants and the overall admitted class for the master’s program to ground their thinking about admissions from a holistic perspective\footnote{1 of the 4 groups did not discuss these questions due to time constraints.}. A common thread in nearly everyone’s answer about what mattered to them in an incoming class was diversity. \anedit{Participants considered a multitude of attributes when defining diversity, often interpreting it in terms of} race and ethnicity (P5, F1, F2), gender (F1, F2), age (P5, F1), socioeconomic background (F1), work background (P5, P6), and general background and experiences (P1, P2, P3, P4, P7, P8, F3). P8 emphasized the importance of a class that mirrored the real world: “Especially since we're going to be working a lot in groups, it would be nice to have like a pretty good representative of, I guess, just the world around us." Student participants also shared a preference for incoming students to have a collaborative nature. As P1 described, he hoped to see “students from different backgrounds who are able to work with people from different backgrounds themselves”, perhaps reflective of coursework or work experiences requiring group work.

While participants mentioned a few quantifiable, academic-related qualities ("education background" -F5, "high grades and GPAs" -F1, "rating of college from which app[licant] has a degree" -F2), the vast majority of applicant attributes were abstract and harder to define or measure. Participants highlighted that applicants needed to exhibit a clear purpose for grad school (P7, F3, F1, P1, P9), be passionate or excited to learn (P2, P3, F3), and be “other-oriented” or service-driven (F2). This overwhelming emphasis on conceptual qualities indicates the nuances in human decision-making and potential challenges a tool may have in assisting or improving human decision-making.

\subsection{Data Exploration and the "All-Features" Model}
Participants next explored historical data of past admissions via the tool and an "All-Features" Model. The tool displayed a subset of 18 features from applications, the distribution of these values, and aggregate data statistics for features with numerical values. We also introduced participants to the feature weights of the "All-Features" Model so that they could discuss their reactions and thoughts about past admissions decision patterns as a group. 

The "All-Features" Model assisted participants in identifying patterns or trends from past decision-making. Students were surprised at some feature weights, such as a small but negative weight for \textit{Awards: Research} and a comparatively strong positive weight for \textit{Work Experience}. P7 observed the weights of the 3 GRE components were all positive and higher than the weight of \textit{Tier of Undergrad Inst.} and shared her surprise and hypothesis for how features may be correlated: “I’m actually really surprised at how much the GRE is taken into account...so I know that the Tier of Undergraduate Institution is kind of largely determined by your, your [socio-economic] class background...so I’m like, I’m wondering if like the GRE weight is meant to like, balance that out.” This point was raised later in the session as another participant shared how they perceived GREs balancing out a low GPA or a low tiered school.

Faculty participants displayed some but less surprise over the weights, perhaps because the majority had previously served on an admissions committee and felt the model was reflective of the patterns they’d observed. But similar to P7, F4 called out the positive weights of GRE features. She suggested an interpretation about what past data may indicate about decision makers: “We’re not using GRE anymore, but it points out the reliance on the GRE in past decisions.” She also observed how these patterns could inform how to advise prospective applicants after seeing the comparatively high weight the model placed on work experience: “It’s almost as if you were to recommend to a student based, again on past [admission’s] experience, how best to prepare for a master’s degree, it would be to get some work experience.”

Participants also generated new ideas for how to \anedit{more effectively} analyze applicant data \anedit{for decision-making} as they assessed the "All-Features" Model. P6 and P9 suggested feature weights could \anedit{be of more use} if broken out by additional criteria such as degree concentration or work experience field. F7 asked whether separate models should be considered for domestic versus international students given the difference in admissions prerequisites for the two (e.g., TOEFL scores for international students). F5 wanted to see information about an applicant’s past institutions in a more comprehensive manner than \textit{Tier of Undergrad Inst.} explaining, “I’d really like to know public schools versus private schools, and size of institution,” expounding later that she values the experiences of students who come from regional or community colleges. 

Though most participants reported basic understandings of programming and AI, reviewing historical data and the "All-Features" Model allowed them to generate ideas and conjectures about past decision-making, which can be used to fine-tune feature engineering. These interactions and ideas suggest that the tool components may support inclusion of participants with expertise or lived experience in the process of participatory AI design.

\subsection{Feature Selection}

Participants sometimes used personal anecdotes and beliefs to explain the inclusion or exclusion of features, highlighting their diverse backgrounds and experiences. P2 explained that since she did not have to submit GRE scores when she applied to grad school, she did not think it was necessary especially when through essays, reviewers could "see their story rather than just a number”. Conversely, P5 included GRE features because as a first generation student, she worked multiple jobs to support herself during school, affecting her GPA, thus “for me, it was like important to submit my [GRE] score so that the admissions committee can see, hey, this person has a 2.8 GPA, but 8 years later...they’ve taken this test and like are a competent reader and math person and writer.” Similarly, F7 shared that growing up in a country that strongly stressed testing had conditioned him to feel favorably towards including GRE scores. 

A handful of participants excluded features based on the limitations they believed an ML model would have when interpreting it. F6 explained that he would personally use descriptive information about awards for making decisions but that the quantified features were insufficient for a model. F5 and F6 agreed on the importance of looking at features like \textit{GPA} and \textit{Tier of Undergrad Inst.} together for context but had different interpretations of how to include features in a model, with F5 wanting to include it while F6 was skeptical whether the specific model would handle features as pairs. P7 omitted \textit{Awards: Arts} after seeing the data and feeling it was inadequate for a model. P3 chose not to include \textit{Gender} based on concerns that the model would leave students out: “The only options there were, were male and female, which leaves out a bunch of people...I don’t know what a model would do if someone said like they were non-binary, like, would it just automatically exclude them because it wasn’t an option in the past data?” 

Some participants wished to update models with custom feature weights to account for different scenarios. F3 suggested that a model handle if-then scenarios when evaluating applicants, where if the value for a feature fell in one range, the model would re-weight another feature in order to balance the first. Some student participants also expressed a desire to customize feature weights. P1 explained wanting features to have "a sliding scale of like how weighted or important it is.” P2 agreed, thinking how applicants with lower GPAs but harder coursework should not be penalized: “Now talking about this a little bit more...if somebody changed their major from like a really hard, you know, engineering major to something else...like it wouldn’t really reflect favorably for them. So I agree with P1, I think I wish that there was some kind of like, weight that it could be put on there.”  

Finally, participants sometimes included features for the nuanced context they felt the features would provide the model. This often arose when discussing \textit{Gender}, \textit{Ethnicity}, and class-related features echoing deliberations earlier over what matters to them in applicants. The feature \textit{First Generation} and the topic of socioeconomic class were openly discussed by faculty and students for inclusion in decision-making. In fact, all but one participant chose to include it in the model. A few student participants explained they viewed \textit{First Generation} as the closest (but "not perfect") proxy to class, sharing concerns of how these applicants face disadvantages around affording school or navigating the application process. \textit{Gender} and \textit{Ethnicity} were approached differently by students and faculty. While both said decisions should not be made solely based on gender or ethnicity, students were more open to discussing the use of them in order to ensure fairness and diversity and contextualize applicants’ experiences, whereas faculty members tended to avoid group discussion over the features in detail. F5 had difficulty articulating why including ethnicity was important for her. A few like F3 wanted to know the features for keeping a check of the diversity of the applicants, but that it otherwise did not factor into her decisions. F6 began sharing that gender and ethnicity should only be taken into account as part of an applicant’s positionality, but discussion over the two features stalled after. We do not suggest that this reflects how faculty members feel about how gender and ethnicity factor into admissions, but instead is more reflective of the rules or expectations that come with such bureaucratic tasks. During interviews, a few expanded on their personal feelings, such as F2 sharing he felt the school had a ways to go in improving ethnic diversity and F4 commenting that historically ethnicity and diversity have been topics of avoidance for admissions.

Though participants' reasoning for feature selection varied from their personal preferences to how they interpreted an ML model, they all contained an undertone in favor of shifting decision-making control (back) to humans. Additionally, their experiences using the tool re-iterate the complicated nature of untangling fairness and equity not just for an automated model but \anedit{within an organization}. That stakeholders have different comfort levels or expectations around deliberating over sensitive features is an important construct to account for in the design of a system.

\subsection{Model Outcome Evaluation}

\subsubsection{Model Performance}
Participants shared their impressions of their model performances and as well as their thoughts on potential harms of models in terms of false positives (i.e., accepting applicants rejected by the past committees) and false negatives (i.e., rejecting applicants accepted by the past committees). 

Although both stakeholder types were aware that the school accepts a limited number of students, in contrast to faculty participants who preferred low false positives, nearly all students shared a preference for models that had high false positives. We observed students displayed inclusion-oriented reasoning, such as P1, P3, and P4 preferring a model that made more false positives, thus “erring on the side of giving people a chance” (P1). P3 felt admitting an “unqualified” candidate would not harm others but could benefit the applicant. Faculty members shared differing opinions. F3 worried about mistakenly admitting a student: “It’s more detrimental to bring a student who cannot succeed. And I think that is, you know, kind of falls on the university.” F5 felt that the pressing potential harm of admissions lies in lack of time or human resources for review, and encouraged a pivot in thinking towards “how do we help committees in the future, as opposed to reinforcing the committees of the past?”

A few participants felt that imperfect accuracy scores could be useful signals to identify areas of decision-making for further investigation. P6 thought an imperfect accuracy could be used to identify whether specific groups of applicants are being unfairly denied: “to find out if there are...certain features of students that fall into that group that humans have denied...that the model accepts or that the humans accept that the model denied.” P9 added that imperfect accuracy scores could be used to improve the model by identifying additional features to include: “if you allow for a gap, we do allow for more of that less specific information. So you can see what was relevant that you need to add into your model in the future.” F3 actually wanted to see a high accuracy on her model, however she shared similar feelings that a gap in accuracy represented an opportunity to improve admissions and the model creation process.

\subsubsection{Personas}

On the Personas screen, participants browsed student profiles and generated hypotheses for the decision-making patterns they were observing. P5 was interested in whether past committees did in fact balance features such as GPAs and GREs. She observed that many applicants with high \textit{GRE Quant} scores and low \textit{GPAs} were accepted in reality. However, after she found a persona with a high GPA/low GRE who was rejected in reality but admitted by the models, she wondered whether reviewers did penalize low scores despite a high corresponding GPA. F2 also noticed a pattern where students accepted in reality but rejected by his models had high \textit{GRE Quant} but lower scores elsewhere, saying “I wonder why we accepted them? Looks like we overvalued the GRE Quant". 

Some explored personas to determine what context to seek out on the rest of an applicant’s package. F1 noticed that the scores of one persona (high GRE, low GPA) resembled someone returning to school from working, and wanted to know more about their work experience background: “Have they been working? Maybe they’ve been working in a library for the last couple of years.” F6 walked through a persona, suggesting possible context for the details he was viewing (“Tier 1 undergraduate institution, so probably a regional schooler”), explaining that his next step would be going into their personal essay for the full story. 

P6 was confused by two personas that the models rejected but the actual decisions were acceptances. Because of the lack of evidence from the features he could see on the screen for accepting them, he was left to conclude that data such as essays made a difference. He felt models could be improved by having access to the reasons behind these candidates’ acceptances: “That would help to understand and improve the model better.” We restate from before that due to privacy issues, we were unable to obtain this data as part of the tool, but we agree with the sentiments of P6 and F6 in the value of the qualitative essays. 

\subsubsection{Participant Criticism and Re-Purposing of Design}
Participant criticism and re-purposing of the web tool aspects often led to proposals of alternate questions around how to support fair decision-making. During model evaluation, we showed participants multiple screens so they could evaluate their results, but observed how some screens were received with counter ideas for how participants wished to use it instead. For example, the Personas screen was constructed to help participants compare the results of the model and past human decision makers at the individual level: it displays acceptance predictions one by one for randomly generated and anonymized applicants. However, as F6 used this, he began critiquing the lack of depth of information it displayed and the use of models for prediction as being "exactly the wrong approach". He suggested he would re-purpose this screen to support a different problem: information summaries to help overwhelmed human reviewers. Students like P1 also re-imagined the screen for a different purpose---rather than using the screen to assess how well the model aligned with past human decisions, P1 referred to the screen as a "bias checker" that reviewers could use the results of to actively challenge their personal biases, an idea reminiscent of how researchers have explored how the disagreements between AI- and human-made decisions can be exploited to improve decision-making overall \cite{de2022doubting, de2020case}. 

\subsubsection{Fairness}
We noticed that the fairness definitions screen of the web tool was challenging for participants to grasp as they struggled during think-alouds and group discussions to share their thoughts. F2 said, “I’ll admit, I didn’t really, yeah this one was a bit above me.” This may have been because though we provided two commonly used mathematical definitions of fairness, equal opportunity and demographic parity may still not have been the most intuitive ways of thinking about an abstract concept as demonstrated by \cite{srivastava2019mathematical}. In addition, definitions of fairness are often not mutually exclusive, as P7 hinted at when pondering how to satisfy individual and aggregate fairness in admissions: "It's hard to tell if the model's fair in the individual...you need it to work [for] an individual and an aggregate." We observed participants found it easier to talk about attributes that should be considered for fair decisions, such as diversity and class rather than discussing statistical measures of fairness, so we shifted discussions around fairness to discuss these attributes instead.

Participants, particularly students, were outspoken about the ways they wanted to see fairness incorporated into decision-making. Students wanted to ensure applicants from traditionally marginalized or underrepresented backgrounds were not disadvantaged. They suggested additional features such as disabilities and veteran status (P2, P3, F4): P3 stated that disabilities can prevent students from having the same opportunities for involvement. Participants insisted on a more reliable indicator for class other than \textit{First Generation} (P3, P5, P7, F2, F3): P7 was emphatic that class cannot be substituted with \textit{First Generation}, and P3 explained that first generation students are disadvantaged when applying for higher education due to lack of parental guidance. Participants also wanted to see improvements on existing features such as making gender more inclusive (P1, P3): P1 wanted gender to include more options than “male” and “female” in order to acknowledge the exclusion that non-binary, gender non-conforming, and transgender people face in opportunities: “It has the potential to really sort of like disrupt somebody...(their) ability to participate fully in school...and all the extracurricular type stuff.” 

\subsection{Post-Session Interviews: Ideas for Future Use of AI or Non-Tech Approaches}

We held wrap-up interviews to get participants' final reflections on their experiences and ideas from using the tool. We wanted to know whether using the tool impacted their perceptions of whether and how AI/ML could be used to assist them. We asked participants to share ideas on how this tool or other AI/ML solutions could be used for admissions. All participants were opposed the use of any fully automated decision-making tool, but some shared how they felt ML could be used to support fair decision-making. Student participants suggested support centered on assisting reviewers in identifying their potential biases. In addition to P1's idea for the web tool as a bias checker, P7 suggested reviewers could use the web tool as a what-if analysis to model and view potential consequences of their human decision-making tendencies before they review applicants. Faculty though, focused on implementations to help reviewers work more effectively and efficiently, such as ML summarizing and describing important characteristics about an applicant to assist reviewers (F6), identifying which applicants were wildcards and the “Larry Birds” of the bunch (F2), and reminding reviewers to pace themselves or review certain components (F5).

We also asked participants to consider non-technological solutions that could apply to a fair admissions process, asking \textit{What non-technology related solutions do you think could help address these [gaps or shortcomings in admissions]?} Their answers converged on the problem \textit{How can the school improve admissions by improving recruitment, enrollment, and performance of diverse and qualified students?} Participants, students in particular, discussed the importance of increasing diversity outside of the review process. They insisted on expanding recruitment and outreach programs to expose a wider audience to the program earlier. P7 emphasized, “if you want more students like me, if you want more Hispanic students”, outreach, mentorship, or pathway programs were necessary. P3 pointed out the importance of downstream efforts as well (i.e., after a student is accepted) to expand funding such that enrollment is not limited to “just those who can afford to pay the tuition”. Their ideas are all striking reminders about how improvements for admissions are not siloed to one area of selections---reviews---but that other parts impact how outcomes ultimately play out. F4 called for more intentional recruitment efforts as well, describing her own experiences of recruiting students of color by building meaningful relationships with prospective students early on. She also urged for colleagues on review committees to hold open conversations about AI, its biases, and the importance of diversity in order to raise awareness, conversations that she was disappointed to note are rarely held. She suggested how a variation of an activity she uses in her teaching, "futuring"---designing futures and imagining what citizens of the future need---could be used by committees to support diversity in practice. Each committee member can come up with scenarios, e.g., a society "coming out of COVID", rate each one's desirability and likelihood, and brainstorm the qualities a person needs to function and thrive, in order to identify diversity or attributes they want to see in future graduate classes. Participants' non-technological ideas echo the sentiments of past research \cite{baumer2011implication,lustig2022designing,hope2019hackathons} to not privilege techno-solutionism.

\section{Discussion}
We designed the \anedit{Deliberating with AI} web tool in order to enable stakeholders to identify ways to improve their decision-making by building an ML model with historic data. Our case study reveals opportunities and challenges for both participatory AI/ML design with stakeholders and deliberation for improving organizational decisions. In this section, we share the insights to inform future research on stakeholder-centered participatory AI design and technology for organizational decision-making.

\subsection{Summary of Results} 
First, we present a brief summary of our findings about how \anedit{Deliberating with AI} helped ground participants' admissions preferences and led to their ideas for improving decision-making. \anedit{Participants used the "All-Features" model and the ML models they created as boundary objects while deliberating with one another. The "All-Features" model allowed them to identify patterns of past decision-making to consider what qualities they cared about in applicants, and their personal ML models helped them discuss admissions contexts and personal experiences to develop a common understanding of what organizational decision-making should entail. Additionally, using ML models led to participants ideating alternative ways for AI/ML to assist reviewers---such as using ML models as bias checkers instead of for predictive capacities---as well as non-technical ideas---such as more intentional recruitment efforts and mentorship programs to identify and support prospective applicants.}

\subsection{Effects of Participatory AI Design in Improving Organizational Decision-Making} 

Using \anedit{Deliberating with AI} to create ML models helped participants generate ideas for future decision-making while also guiding them in a process of participatory AI design. Below, we describe how the web tool helped participants execute these aims, specifically the role of deliberating with ML models as boundary objects to help participants recognize shared interests and the complexities of decision-making, and the web tool itself as an applied method for prototyping participatory AI design. We offer ideas for how future research endeavors may be able to use these in support of organizational decision-making and participatory AI design.

\subsubsection{Deliberating with ML Models as Boundary Objects to Uncover Opportunities for Organizational Decision-Making}

Boundary objects are defined by \citet{star1989structure} as objects that are "both plastic enough to adapt to local needs and constraints of the several parties employing them, yet robust enough to maintain a common identity across sites" and further adapted in design and cooperative work research \cite{lee2007boundary,john2004identifying,cai2021onboarding} to be "common frames of reference that enable interdisciplinary teams to cooperate" \cite{cai2021onboarding}. Given these descriptions, we propose that participants' individual models can be viewed as boundary objects: they act as frames of reference that were used during deliberations by participants to convey their own rationale and understand other people's reasoning. This in turn led to their unique and varied ideas for how to support organizational decision-making, from technical solutions to mitigate reviewer biases to non-technical approaches to support diverse recruitment.

As boundary objects, ML models gave participants a way to reach a common understanding of the beliefs they shared and the complexity of their differences within the problem space. With these models, participants had an accessible starting point for deliberation because all the ML models held a common identity for them to grasp---the features used in them are recognizable attributes usually found on applications. Then as deliberation progressed, the discussions were enriched because ML models and features took on localized meanings based on each participant's lived experiences---e.g., F1 was sympathetic to including GRE scores because she believed her high GRE score secured her graduate school acceptance, balancing her low GPA. Talking about their models allowed participants to convey and listen to these types of anecdotes which was invaluable for participants like F5: "I learned so much from my colleagues, especially when they give their own personal experiences." Finally, we observed that deliberating with ML models ultimately supported participants in critically assessing admissions as a whole and surfacing how to improve future organizational decision-making. F4 shared that discussions around feature selection and the "All-Features" Model made her think beyond how to assess applicants and instead how to expand opportunities to support diverse candidates (“how can we decolonize admissions?”). This also offers initial support that our web tool can be useful for problem formulation in AI/ML \cite{passi2019problem}. While our web tool did not have participants begin with AI/ML problem formulation as our goal was to explore ways to improve future decision-making in general without requiring AI/ML processes, participants shared throughout the session and wrap-up interviews how using the tool made them think of alternate problems to consider as well as ideas and preferences of technical solutions (e.g., the use of ML for descriptive information about candidates -F6) vs. non-technical solutions (e.g., investing in outreach and pathway programs to expose more students of color to the program and increase diverse recruits -P7).

In designing \anedit{Deliberating with AI}, one of our reasons for users creating ML models was to help support them in structured discussion. We were not sure exactly how users would incorporate ML models in group deliberation, but we imagined that models would play a role in shaping how users organize their thoughts or share perspectives on decision-making. It makes sense then, that participants naturally used ML models as boundary objects when deliberating because their ML model acted as a mechanism for them to contextualize their feature selections and views. This leads us to wonder what other ways researchers may be able to explore deliberation with ML models as boundary objects in the future. In our case study, most participants had basic understandings of AI and computational models, but using ML models as boundary objects helped them frame what they wanted AI systems to reflect. We held separate sessions for decision makers and decision subjects, but often participatory AI design requires input of diverse stakeholders and the collaboration of stakeholders with designers and \anedit{data scientists}. Thus it may be beneficial for researchers to conscientiously explore with groups of mixed disciplines (e.g., stakeholders, designers, \anedit{data scientists}) or mixed stakeholders (e.g., decision makers, decision subjects) whether ML models can be used as boundary objects to share expertise and align on AI design. Additionally, researchers may also study how ML models as boundary objects can be formed more robustly to support stakeholders communicating to technologists (e.g., \anedit{data scientists}, designers, practitioners) the complex nuances of organizational decision-making to take into account in an AI/ML-based solution.

\subsubsection{Using Deliberating with AI as a Prototyping Tool for Centering Stakeholders in Participatory AI Design}
Participatory AI design has thus far explored how to create AI models with individuals or groups. But often stakeholders involved are non-ML experts who may struggle with understanding how their decisions, such as feature selections, actually impact the outcomes of the model. To support non-ML experts, researchers have explored ways to make ML understandable, such as designing visualizations to convey model trade-offs to users \cite{ye2021wikipedia, shen2020designing}. 

We propose \anedit{Deliberating with AI} as another method, a prototyping tool for researchers to use in advancing stakeholder-centered participatory AI design. Based on the feedback from participants about their experiences using the tool, it became evident that one of the biggest benefits they saw in creating ML models was being able to ground their ideas in practice, and then immediately view and explore the results. By engaging in hands-on practice with models, participants were able to uncover ideas for how ML can assist decision-making responsibly or ways it fell short. For example, for P7, the hands-on practice of reviewing personas helped her realize that some features she valued abstractly she did not care about in practice: "When I'm sitting in the chair, looking at it, it's like, I think awards are completely irrelevant." For P1 and P3, comparing their individual and group models’ feature weights let them see how a model interpreted features they were unsure about earlier. P3 decided she would have included \textit{First Generation} in her individual model after seeing in the group model that including it did not penalize those students from being admitted. The differences in people's expectations and realities of their models, as well as the changes they often wanted to make after reviewing models mark how a prototyping tool for ML models can inform and empower participants as they engage in AI design. 

We contend that research intended for advancing stakeholder-centered participatory AI design should not only involve stakeholders, but help them feel empowered in the choices they make and even curious about how things work. We were reminded of this after a comment by P7 who, after hesitating over a feature selection, decided to include it simply out of curiosity, exclaiming "this is just like a first draft!" We agree: an ML model does not have to be perfect to allow participants to engage with it and surface ways to improve decision practices. We encourage researchers to explore the use of prototyping tools or the design of other participatory AI aids to empower stakeholders throughout their involvement of AI design. 

\anedit{In pursuit of this, we highlight two directions for future work on participatory AI. First, work expanding participation of non-ML experts in AI design can focus specifically on data exploration and feature selection. We observed that in these stages, reflecting and deliberating over data and features was more approachable for participants as they could use their ML models as boundary objects to discuss applicant features anecdotally. These stages may present fewer barriers for engaging with non-ML stakeholders who are novices in data analysis or technology. 

As a second line for future work, we suggest that researchers investigate additional methods to improve reflection and deliberation at stages such as model training and evaluation which may be less intuitive to non-ML experts. For example, model evaluation introduced metrics to assess models such as precision and recall which are not common knowledge for everyone and may have stilted reflection and deliberation. In our study, the Personas tool allowed participants to browse model outcomes and concretize abstract preferences. Future work can explore alternative approaches that similarly facilitate reflection and deliberation in non-traditional ways during model training and evaluation.}

\subsection{Considerations When Working with Different Stakeholder Types}
Participatory methods often emphasize the importance of engaging directly with impacted stakeholders in the design of (automated) systems or processes \cite{delgado2021stakeholder}. However, guidelines or considerations for engaging with different types of stakeholders are not always provided, although some resources share considerations for working with impacted stakeholders such as \citet{nelson2020toolkit} providing suggestions for navigating power dynamics and equity awareness and \citet{harrington2019deconstructing} sharing principles for equitable participatory design when working with underserved populations. We explore questions that emerged as we engaged with two stakeholder types---decision makers and decision subjects---and initial ideas for addressing these questions. 

\subsubsection{Where Decision Makers and Decision Subjects Converge and Diverge}
Faculty (decision makers) and students (decision subjects) exhibited similarities and differences in their views of certain features and how decision-making should be conducted. Early on, both faculty and students expressed a desire for decision outcomes to reflect diversity, and this often emerged through group discussions and the ideas they came up with around \anedit{improving future decision-making}. Both faculty and students also used similar logic during feature selection, choosing to include features based on personal experiences or how they hoped a model would use it.

However, faculty and students differed in a few ways such as the error types they believed decision-making should aim to minimize. Both students and faculty recognized the competitive nature of admissions and were aware that there is a limit to how many students the school can accept. On one hand, students overwhelmingly believed that false positives (i.e., accepting applicants rejected by the past committees) were less harmful while faculty members felt the opposite. Students viewed the former as giving “people a chance”, with P1 adding that he felt false positives could be a way of “correcting for the bias in the data” given potential historical biases. Conversely, faculty believed false positives were more harmful for schools: F3 explained that if an accepted student was not successful, the institution would be at fault for not providing adequate support. Another factor for why faculty and students differ may be attributed to how faculty currently lack measurements to assess whether accepted students were successful and trust their colleagues who made historical decisions. The school does not currently track information for all students that can be used for assessing their success after being accepted such as employment information after graduating, and graduate GPAs are commonly skewed higher and therefore can't be used as a differentiating variable to measure success. Differences in temporal proximity to applying to grad school themselves may also explain the divergence \cite{liberman2007psychological}---in that sense, it may be harder for faculty to conceptualize potential harms of admission "misses" while students have recently been applicants and can recall more viscerally being accepted or rejected. 

\subsubsection{How to Accommodate Differences Between Decision Makers and Decision Subjects}
Differences in how decision makers and decision subjects view potential harms and how to use resources to improve decision-making present unique challenges for the systems that support deliberation. 

First, the difference in how faculty and students viewed harms of admitting someone who is "unqualified" and rejecting someone who is "qualified" signifies a consequential difference how they might view a new admissions policy should be designed. Faculty took on the perspective of institutions in how "mistakes" may impact the school while students took the perspective of applicants and the impacts on them. This difference in perspective around harm translates to perceptions of fairness and trustworthiness of a decision process. It raises questions around how these differences should be accounted for in the creation of AI/ML models or decision practices to be considered trustworthy by all stakeholder types. 

Second, faculty and students presented different ideas for how technological or non-technological tools can help, which may indicate a potential tension over how resources should be used within various problem domains. For example, as related to non-technological support for selection processes such as admissions and hiring: is it more important to expend money and effort to ensure future recruits come from diverse communities? Or is it more important to devote resources to the immediate hiring and training of reviewers? Resolving this difference requires asking another question: how should the input of both decision makers and decision subjects be combined so both feel supported in their goals, especially as stakeholders may take on different roles in their lifetime (e.g., changing from decision subjects to decision makers or decision influencers)? 

Some ideas for how designers can explore this include holding public deliberations with different stakeholder types to share each other's outlooks \cite{fung2007minipublics} and/or using a voting system like WeBuildAI to weight each stakeholder type's input for algorithm design \cite{lee2019webuildai}. But whether the outcome is to equally weight each person's input, place higher weight on a certain stakeholder's input, or another method, the decision should involve representative stakeholder groups for each specific situation. To support stakeholders then, a focus for designers may be assisting them in how to recognize where their agreements lie and how to frame their arguments with one another. 

\subsection{Importance of Care in Design: Working With Data and Navigating Power Dynamics} 
Although all of our participants were able to complete the session and create an ML model, we stress the importance of care when having users work with data. While the methods of data exploration and deliberation that we used were able to support our participants in making sense of past data, working with data is still a daunting task for many and intimidating to attempt alone and/or in front of others. 

Some of our participants shared insecurities about exploring data due to their background or session nerves. In these instances, facilitators worked to ameliorate apprehensions by breaking down concepts or focusing on qualitative questions. P5 shared, “I’m really not familiar with working with datasets at all, so I don’t know if I’ll make any intelligent decisions.” She began to enjoy using the web tool though, particularly the Personas screen, exclaiming as breakout rooms closed, “I want to keep working!” F2 indicated frustration at the start of Model Evaluation, explaining, "I’m not 100\% sure I’m going to be able to figure it out.” With facilitator guidance, he began to gain confidence as he explored the various screens. However, facilitator assistance was not always enough: F1 declined to explore parts of the tool, saying, “That makes me feel stupid”, and sharing during the interview that, “This could have been like an all day thing, and I still probably wouldn't have had enough time to figure it out.”

Additionally, the topic of admissions may be unsettling for participants who feel uncomfortable sharing their personal views in groups or with strangers or have negative past experiences. We attempted to reduce negative associations for participants such as interjecting when needed, redirecting questions that stirred anxiety, and reassuring them that there was no right or wrong answer. It may not be obvious though, during group discussions, if there is a discomfort. F4 shared in their debrief the hurt she felt when hearing a colleague explain their inclusion of GREs. She felt this privileged a Eurocentrist view in admissions, unfairly disadvantaging groups like Indigenous tribes who have different testing cultures and low access to test preparation resources: “I did feel very, actually had to look away from the screen...that was very hurtful for me to hear that because it tells me that we should require...a Blackfeet person living in a rural area in Montana to do what we did.” She also commented about the power dynamics that existed in the room of participants given varying experience in admissions, as a faculty member, and tenure at the school. This within-group power dynamic surprised us as we had previously only been cognizant of ensuring students and faculty did not overlap given that asymmetry of power, but is another consideration for how to responsibly engage with communities and different stakeholder types in the design of participatory algorithms. 

In some instances, it may be possible to keep identities anonymous to temper power dynamics, where strangers are brought together or where participation is through text mediums. Amongst colleagues and when using video or voice formats, anonymity can be integrated by using forums for follow-ups where pseudonyms can be used. Related research that has emphasized care in design has suggested circumventing issues through prioritizing specific voices \cite{hope2019hackathons}, building rapport with groups beforehand to support trust \cite{le2015strangers}, and evaluating session materials for how they may privilege certain groups of people \cite{harrington2019deconstructing}. An additional idea to explore is how to "artificially" establish common ground to encourage sharing openly by pairing participants based on similar responses and later rotating and widening groups to share diverse perspectives. \citet{burkhalter2002conceptual} suggests that establishing common ground in anticipation of deliberation can help with participation---designing so that participants perceive common ground could be useful with helping people of different backgrounds gain familiarity and comfort.

Even so, addressing participant comfort of analyzing or using data in a session remains a challenge. Though we see merit in pursuing how to make data accessible to non-ML experts through visualizations \cite{ye2021wikipedia, shen2020designing}, we suggest a different question for researchers to consider: how can we leverage each person's unique strengths in the process of analyzing data? \cite{williams2014city} and \cite{d2017creative} are two examples encouraging creative approaches for exploring and understanding data, for example by using physical space and movement to engage with data \cite{williams2014city} or having users partake in "data biographies"---(re-)contextualizing data by qualitatively investigating its origins and purpose \cite{d2017creative}. Designers may consider how to construct tools or sessions that cater towards both those with an affinity towards data analysis vs. those who feel most comfortable in other modes of inquiry or contribution (e.g., writing white papers or policies that use the insights of data analytics by the former group).
 
\section{Limitations and Future Work}
We acknowledge limitations of our study. Our tool used only quantitative data due to the privacy issues associated with qualitative data. We set sessions to 2-2.5 hours due to the challenges of scheduling and recognizing that longer sessions would lead to more fatigued participants. This limited the time for participants to digest and fully explore the information on the tool. Splitting the session into a series can be one way to allow participants to gradually build knowledge and comfort using models in future studies. 

\anedit{Clean datasets are not always readily available (as our team also experienced during this process), introducing an additional consideration for future users of the tool.} Due to the unknown variance in baseline ML knowledge of our participants and the explainability of linear regression algorithms, we used linear regression to minimize cognitive load on participants; however, catering our explanations to those with little ML knowledge may have made the tool and model overly simplistic to participants with greater ML experience and limited what they were able to contribute. Also because of the limited time, we had to omit more sophisticated designs such as allowing a non-linear or iterative use of the tool, as well as having participants deliberate over and select which ML algorithm to use for model training which was mentioned previously. Future studies should explore iterative use of the tool---people changing features, retraining, and examining the models. Finally, we were only able to recruit from students who were accepted and enrolled: future studies may be done with those who have not yet but plan to apply, applied and were rejected, and were accepted but chose not to come. We intentionally held sessions for only faculty members or only students due to the inherent power dynamics between the two. Future work can explore ways to account for power dynamics in deliberation as mixing stakeholder types may affect different deliberation outcomes and group models.

\section{Conclusion}
Many important lines of research are working towards ensuring algorithmic decision-making systems are created to be equitable and fair. While these approaches focus on how to improve or create a machine learning or automated system, we propose using machine learning to improve human decision-making. We created a web tool, \anedit{Deliberating with AI}, to explore how to support stakeholders in identifying ways to improve their decision practice by building an ML model with historic data; we demonstrate how faculty and students used the web tool to externalize their preferences in ML models and then used their ML models as boundary objects to share perspectives around admissions and gain ideas to improve organizational decision-making. We share the insights our case study to inform future research on stakeholder-centered participatory AI design and technology for organizational decision-making.

\section{Acknowledgements}
We wish to thank Bianca Talabis for creating the illustrations in our paper, the anonymous reviewers whose suggestions greatly improved our paper, and our participants for their thoughtful engagement and insights that they shared with us during the sessions. This research was partially supported by the following: the National Science Foundation CNS-1952085, IIS-1939606, DGE-2125858 grants; Good Systems, a UT Austin Grand Challenge for developing responsible AI technologies\footnote{https://goodsystems.utexas.edu}; and UT Austin’s School of Information. 

\bibliographystyle{ACM-Reference-Format}
\bibliography{references.bib}

\section{Appendices}

\begin{table*}
\resizebox{0.8\linewidth}{!}{%
\begin{tabularx}{\linewidth}
{b{0.25\textwidth} 
b{0.7\textwidth}} 
\toprule
\textbf{Feature Names}
& \textbf{Description}
  \\\toprule

GRE Verbal \%, GRE Quant \%, GRE Analytical \% 
& Directly mapped from GRE percentile columns in the dataset.
\\ \toprule
Tier of Undergrad Inst.
& Mapped by taking the undergraduate institution the applicant most recently matriculated from to a tier. 4 is the highest tier, while 1 is the lowest. Tiers were formed by aggregating National, Regional, Country, and International school ranking lists from US News. 
\\ \toprule
GPA 
& Mapped from the GRE column in the dataset that the graduate school calculated from an applicant's upper level classes.
\\ \toprule
Master's Held, Doctorate Held, Special Degree Held
&  Mapped as a binary from whether the applicant reported obtaining one of these degrees in their education history.
\\ \toprule
Awards: Arts, Competition, Leadership, Research, Scholastic, Service  
& Formed by hand-coding the 3 free-form text fields each applicant could use to report honors/awards into categories and summing these categories for each applicant.
\\ \toprule                                           
Gender
& Mapped directly from the gender column of the dataset, historically limited to only male or female.
\\ \toprule
Ethnicity
& Mapped from the primary ethnicity column of the dataset.
\\ \toprule
First Generation
& Formed by the education history columns of the applicant's parents/guardians. If all reported history is below a Bachelor's Degree, this value is Yes.
\\ \toprule
Work Experience
& Calculated from the applicant's reported work history dates, by subtracting their earliest work start date from their most recent end date.
\\ \toprule
\end{tabularx}
}
\captionsetup{width=0.7\textwidth}

\caption{Logic for how the 18 features were formed from the original dataset.}
\label{table:TableOfFeatureLogic2}
\end{table*}

\begin{table*}
\resizebox{0.7\linewidth}{!}{%
\begin{tabularx}{\linewidth}
{b{0.25\textwidth} 
b{0.5\textwidth} 
b{0.075\textwidth} 
b{0.075\textwidth}}
\toprule
\textbf{Feature Name}
& \textbf{Description}
& \textbf{Incl. by Students}
& \textbf{Incl. by Faculty}
  \\\midrule

GRE Verbal \% 
& Percentile of the applicant’s GRE Verbal score.
& 78\% & 57\%                                                \\ \midrule
GRE Quant \% 
& Percentile of the applicant’s GRE Quantitative score.
& 67\% & 57\%                                                \\ \midrule
GRE Analytical \%
& Percentile of the applicant’s GRE Analytic (Writing) score.
& 67\% & 57\%                                                \\ \midrule
Tier of Undergrad Inst.
& Tier 4 being top institutions and 1 being bottom. Primarily determined by aggregating several US News rankings.
& 44\% & 86\%                                                \\ \midrule
GPA 
& Upper level grade point average of the applicant, as calculated by their grades in senior level courses (e.g., typically taken in their third year and beyond if considering their bachelor’s experience)
& 89\% & 86\%                                                \\ \midrule
Master's Held
& Whether the applicant holds a master’s degree
& 11\% & 86\%                                                \\ \midrule
Doctorate Held
& Whether the applicant holds a doctorate degree
& 11\% & 43\%                                                \\ \midrule
Special Degree Held
& Whether the applicant holds a special degree
& 11\% & 43\%                                                \\ \midrule
Awards: Arts  
& The number of awards or honors that the applicant listed, related to the arts (e.g., creative writing, English, music, etc.)
& 78\% & 43\%                                                \\ \midrule
Awards: Scholastic
& The number of awards or honors that the applicant listed, related to receiving scholarships or being holding top student rankings such as valedictorian.
& 100\% & 86\%                                                \\ \midrule
Awards: Research
& The number of awards or honors that the applicant listed, related to research experience such as independent study, research grants, or writing a thesis.
& 89\% & 86\%                                                \\ \midrule
Awards: Service
& The number of awards or honors that the applicant listed, related to service or volunteering.
& 67\% & 71\%                                                \\ \midrule
Awards: Leadership
& The number of awards or honors that the applicant listed, related to holding leadership positions.
& 78\% & 57\%                                                \\ \midrule
Awards: Competition
& The number of awards or honors that the applicant listed, related to competitions or contests relevant to academia (e.g., creative writing, English, music, etc.)
& 78\% & 71\%                                                \\ \midrule
Gender 
& Applicant’s self-reported gender. Historically, this has been limited to two choices, male or female.
& 11\% & 57\%                                                \\ \midrule
Ethnicity
& Applicant’s self-reported ethnicity. Historically, this has been limited to five race/ethnicity categories.
& 44\% & 71\%                                                \\ \midrule
First Generation
& Whether the applicant is the first in their family to receive a bachelor's degree. Inferred by the education level of an applicant’s parent(s).
& 89\% & 100\%                                                \\ \midrule
Work Experience
& How many years an applicant has worked, defined by the time between the applicant’s earliest work start date and most recent work end date.
& 100\% & 100\%                                                
\\ \bottomrule
\end{tabularx}
}
\captionsetup{width=0.7\textwidth}

\caption{Set of 18 features displayed in the tool during Feature Selection as well as in the AllFeaturesModel.}
\label{table:TableOfFeatures2}
\end{table*}
\clearpage
\received{July 2022}
\received[revised]{October 2022}
\received[accepted]{January 2023}

\end{document}